\documentclass[12pt]{article}

\usepackage[T1]{fontenc}

\usepackage{color}
\usepackage{colordvi}
\usepackage{amsmath}
\usepackage{amssymb}
\usepackage{amsfonts}
\usepackage{graphicx}
\usepackage{multirow}
\usepackage[dvipsnames]{xcolor}
\usepackage[colorlinks=true,urlcolor=blue,linkcolor=blue,citecolor=blue]{hyperref}
\usepackage{cite}
\usepackage{cancel}
\usepackage{booktabs}
\usepackage{caption}
\usepackage{subcaption}
\usepackage{float}

\usepackage[top=1.135in,bottom=1.17in,left=1.16in,right=1.16in]{geometry}
\numberwithin{equation}{section}

\allowdisplaybreaks

\newcommand{\eq}[1]{\begin{align}#1\end{align}}

\begin{document}
\thispagestyle{empty}

\begin{center}

{\bf\Large \boldmath Resonant Majorana neutrino effects in $\Delta$L=2 four-body hyperon decays}

\vspace{50pt}

Gerardo Hern\'andez-Tom\'e$^{\dagger}$, Diego Portillo-S\'anchez$^\star$,  and Genaro Toledo$^{\dagger}$ 

\vspace{16pt}
{\small
{\it $\dagger$  
Instituto de F\'isica, Universidad Nacional Aut\'onoma de M\'exico, AP 20-364, Ciudad de M\'exico 01000, M\'exico
} \\
{\it $\star$Departamento de F\'isica, Centro de Investigaci\'on y de Estudios Avanzados del Instituto Polit\'ecnico Nacional} \\
{\it Apartado Postal 14-740, 07000 M\'exico D.F., M\'exico}
}
\vspace{16pt}


\today

\vspace{30pt}

\end{center}

\begin{abstract} 
We computed the $\Sigma^{-}\to n\pi^+e^-\ell^-$ ($\ell=e,\mu$),  $\Xi^-\to \Lambda\pi^+e^-e^-$, and $\Lambda\to p\pi^+e^-e^-$ lepton number violating (LNV) hyperon decays mediated by a resonant Majorana neutrino. The expected hyperon production rate of experiments like BES-III of around $10^6-10^8$ may allow searching for these rare hyperon decays at enough sensitivities. We illustrate the limits on the new heavy mixing parameters derived from these hyperon channels and compare them with other LNV meson decays in similar mass regions of the resonant neutrino state.
\end{abstract}


\section{Introduction}


The study of hyperon decay properties had a golden era some sixty years ago when Cabibbo proposed the universality of charged weak interactions in semileptonic decays \cite{Cabibbo:1963yz}. 
Hyperon semileptonic decays were used to measure the weak charges in strangeness-changing transitions and to extract the Cabibbo angle $\sin \theta_c$. On the other hand,  non-leptonic decays allowed to measure the hyperon polarizations and to determine the final state interactions phases \cite{Hara:1966qjc, Bunce:1976yb, LeYaouanc:1978ef}. The field of hyperon physics was somehow abandoned with the advent of high intensity kaon beams which allowed to extract the Cabibbo angle with reduced strong interactions uncertainties. Until the late nineties, only a few searches of rare and forbidden hyperon decays were reported \cite{Littenberg:1991rd}.  In the last twenty-five years, a few more data on allowed, rare and forbidden hyperon decays were reported by the HyperCP \cite{HyperCP:2005sby}, NA48 \cite{Raggi:2015noa}, LHCb \cite{LHCb:2017rdd}, KTeV \cite{KTeVE832E799:1999tte}, BESIII \cite{Li:2016tlt, BESIII:2020iwk} collaborations.

The BESIII hyperon physics program has brought a renewed interest in this field thanks to the large dataset of baryon-antibaryon pairs produced in $J/\psi$ and $\psi(2S)$ decays \cite{Li:2016tlt}. Owing to the non-negligible branching fractions for these decays, the large production rate of these charmonium states would allow the production of $10^6-10^8$ hyperon pairs of different species. This opens the possibility of improving measurements of allowed and rare hyperon decays that will set strong limits, for example, on the rare FCNC hyperon decays with charged lepton or neutrinos pairs \cite{Li:2016tlt}. Similarly, searches for forbidden (lepton number or baryon number) decays can be pursued, allowing to test models that include the violation of these accidental symmetries \cite{Li:2016tlt}. 

At present, the observation of neutrino oscillation represents one of the most thrilling discoveries in particle physics, setting new questions about the nature and origin of their tiny masses. The most promissing approach to establish if neutrinos are their own antiparticles is to search for lepton number violating processes ($\Delta$L=2) which would be only possible if that is the case. Despite the neutrinoless double beta decays in nuclei ($0\nu\beta\beta$) are the most extensively studied and promising laboratory to give an answer on this matter, alternative and complementary searches for other LNV processes can play also a crucial role in current and future experiments, since they provide information on specific energy windows.

In this paper we focus on the  $B_A(p_A)\to B_B(p_B)\ell_1^-(p_1)\ell_2^-(p_2)\pi^+(p_\pi)$  lepton number violating (LNV) decays ($B_{A,B}$ denote hyperon states, see Fig. \ref{diagram}). Specifically, we will consider the following channels: $\Sigma^{-}\to n\pi^+e^-\ell^-$ ($\ell=e,\mu$),  $\Xi^-\to \Lambda\pi^+e^-e^-$, and $\Lambda\to p\pi^+e^-e^-$. This kind of decays have not been studied before, and they can be induced by the resonant enhancement of intermediate mass Majorana neutrinos\footnote{These novel channels extend the search of similar LNV effects performed in semileptonic baryon, meson and tau decays \cite{Atre:2009rg, Abada:2017jjx, LopezCastro:2012udb, Castro:2013jsn, Milanes:2016rzr, Das:2021kzi, Das:2021prm, Zhang:2021wjj, Cvetic:2016fbv, Cvetic:2017vwl, Cai:2017mow, Mejia-Guisao:2017gqp, Yuan:2017uyq, Li:2018pag, Milanes:2018aku, Cvetic:2019shl, Chun:2019nwi, AlvesJunior:2018ldo}.}. LNV hyperon decays of the form $B_A^- \to B_B^+\ell^-\ell'^-$ have been studied before in Refs. \cite{Barbero:2002wm, Barbero:2007zm, Barbero:2013fc, Hernandez-Tome:2021byt}. These processes are mediated by a virtual Majorana neutrino and are similar to neutrinoless double beta decays. On the other hand, resonant production of Majorana neutrinos are possible for a limited range of their masses in such a way that they can be produced on their mass-shell. Contrary to production of virtual Majorana neutrinos processes with rates of $O(G_F^4)$, the rates for production of resonant Majorana neutrinos becomes of $O(G_F^2)$ \cite{Atre:2009rg, Abada:2017jjx}, which allows to place better constraints of their parameter space even with upper limits given by current experimental sensitivities. 

In the following we present the formalism to describe these processes and introduce the integration method for four-body decays, which extend the one followed in the 3-body case \cite{Atre:2009rg, Abada:2017jjx} and allows to properly account for the different charged leptons flavor case.
Given the clean experimental signature, one may expect that very strong limits can be set on the branching fractions of these decays, similar to existing limits on other $\Delta$L=2 meson decays. Therefore, it would be interesting to explore if similar limits on the parameter space of resonant Majorana neutrinos can be obtained from the proposed four body hyperon decays.

\section{Computation}\label{S-com}

\begin{figure}[]
\begin{center}
\begin{tabular}{cc}
\includegraphics[scale=.6]{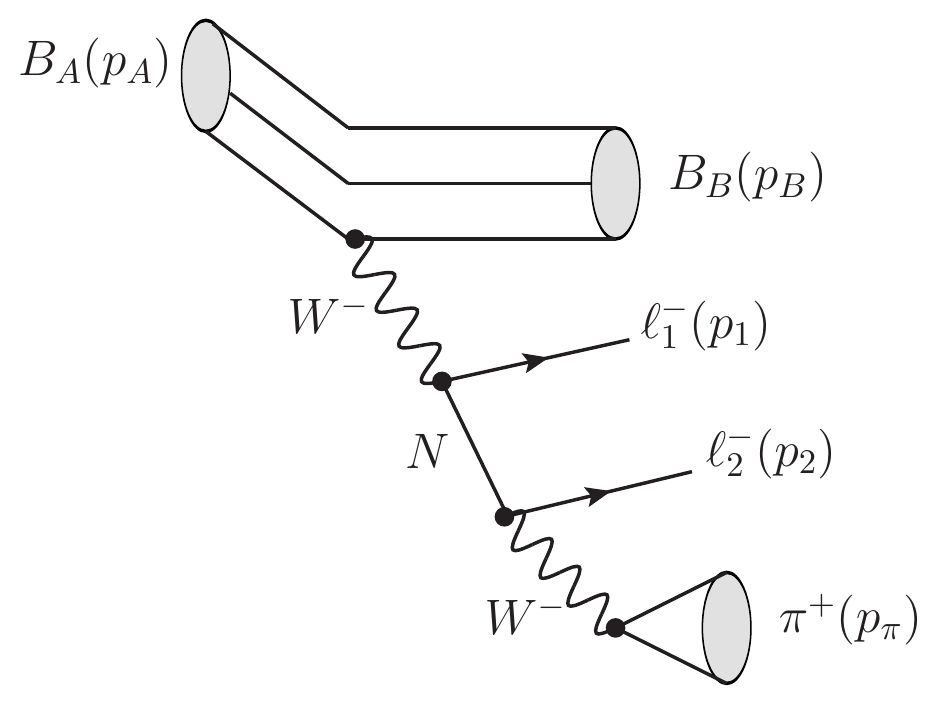}&\includegraphics[scale=.6]{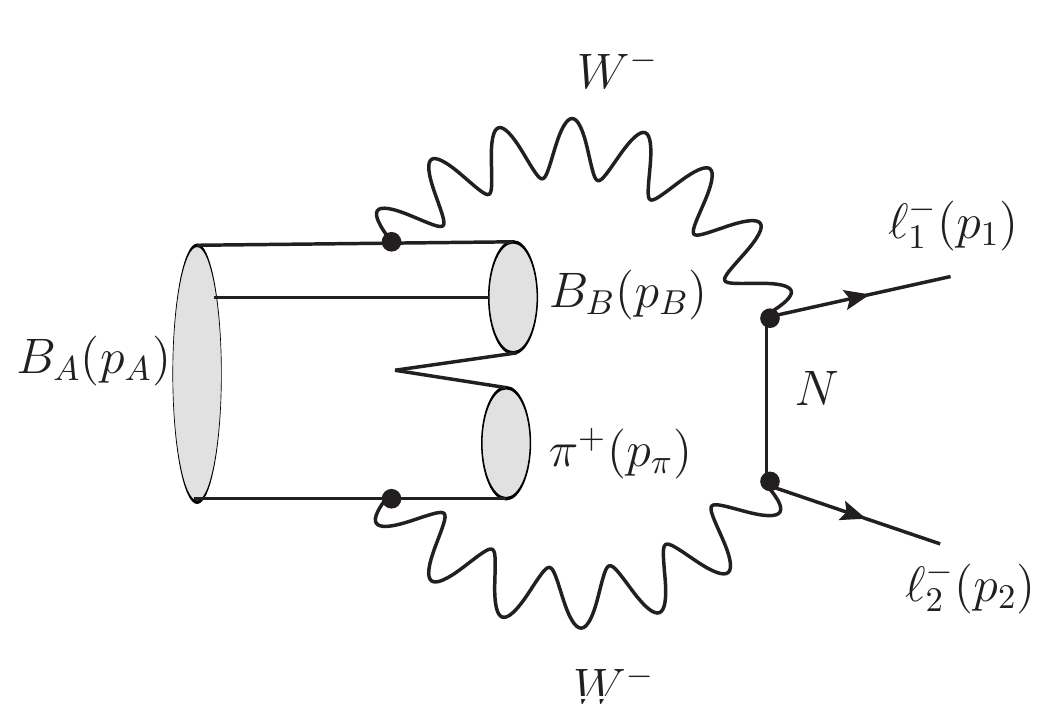}\\ (a) & (b)
\end{tabular}
\caption{Feynman diagrams for the four-body $\Delta L$=2 hyperon decays mediated by a resonant heavy Majorana neutrino $N$.  We consider the following channels: $\Sigma^{-}\to n\pi^+e^-\ell^-$ ($\ell=e,\mu$),  $\Xi^-\to \Lambda\pi^+e^-e^-$, and $\Lambda\to p\pi^+e^-e^-$. Note that diagram (a) is the dominant one when the neutrino is on-shell because its contribution is enhanced due to a resonance effect, opposite to diagram (b) where the neutrino can not become into a resonant state.}
\label{diagram}
\end{center}
\end{figure}

Adopting the convention for the neutrino states on Ref. \cite{Atre:2009rg}, let us consider an scenario where the leptonic sector incorporates a number $n$ of singlet right-handed neutrinos $N_{R_j}$ ($j=1,2,\ldots n$) in addition to the usual three left-handed $SU(2)$ lepton doublets $L^T_{iL}=(\nu_i, \ell_i)_L$. In such scenario, after the proper mass matrix diagonalization, the charged lepton current relevant for our computation can be written as follows
\eq{
\mathcal{L}_{W}=-\frac{g}{\sqrt{2}}W^+\bigg( \sum_{\ell=e,\mu,\tau}\sum_{i=1}^3U_{\ell i}^*\bar{\nu_i}\gamma_\mu P_L\ell+ \sum_{\ell=e,\mu,\tau}\sum_{j=4}^{3+n}V_{\ell j}^*\bar{N}_j^c\gamma_\mu P_L \ell\bigg)+\textrm{h.c.},
}where $P_L=(1-\gamma_5)/2$ is the left-handed chirality projector, $N^c=C\bar{N}^T$ is the charge conjugate spinor, and $U_{\ell j}$ ($V_{\ell j}$) describes the lepton mixing matrix elements for the light (heavy) neutrino states. 

Similar to previous works, we base our analysis considering the case of a simply minimal scenario with only one heavy Majorana neutrino $N$, with the corresponding mass $m_N$ and mixing with the charged lepton flavor $V_{\ell N}$ ($\ell=e,\mu,\tau$) \footnote{This minimal scenario is not able to explain the current data coming from neutrino oscillations experiments but represents a simple approach to encode the effects of a larger number of heavy states present in well-justified massive neutrino models. Recently, the interference effects in extensions with at least two heavy neutrino states for three-body meson LNV decays have been reported in \cite{Abada:2019bac, Godbole:2020jqw, Zhang:2020hwj}.}. The relevant diagram for the $ B_A(p_A)\to B_B(p_B)\ell_1^-(p_1)\ell_2^-(p_2)\pi^+(p_\pi)$ LNV hyperon decays is depicted in Fig. \ref{diagram}(a), and its amplitude can be written as follows
\eq{
\mathcal{M}_{1}=\left(\frac{G\,V_{\ell_1 N}V_{\ell_2 N}f_\pi m_N}{a_{1}+i\Gamma_N m_N}\right)\ell_{\mu\nu}(p_{1},p_{2})\, H^\mu(p_B, p_A)\, p_\pi^\nu,\label{amp-com-1}
}where $a_{1}\equiv(p_A-p_B-p_{1})^2-m_N^2$, and $p_A-p_B-p_{1}=p_\pi+p_{2}$ is the momentum carried out by the heavy neutrino $N$, and we have defined $G\equiv G_F^2V_{us}V_{ud}$. The leptonic and hadronic parts are given by
\eq{
\ell_{\mu\nu}(p_1,p_2)&\equiv \bar{u}(p_1)\gamma_\mu \gamma_\nu(1+\gamma_5)v(p_2),\\
H^\mu(p_B,p_A)&\equiv \langle B_B(p_B) \vert J_\mu  \vert B_A(p_A)\rangle.\label{Hmu}
}The hadronic current $J_\mu$ is parametrized in terms of six form factors which are determined from the well-known lepton number conserving  hyperon decays $B_A\to B_B \ell^{-} \bar{\nu}_{\ell}$  ($\ell=e,\, \mu $) \cite{Garcia,Schlumpf:1994fb,Ratcliffe:2004jt, Mateu:2005wi}:
\eq{
\langle B_B(p_B)\vert  J_\mu \vert B_A(p_A) \rangle &= \bar{u}(p_B)\bigg[ f_1(q^2)\gamma_\mu +i f_2(q^2) \frac{\sigma_{\mu\nu}q^\nu}{M_A}+\frac{q_\mu f_3(q^2)}{M_A}\label{ff}\\
&\quad\quad\quad\,\,\,+ g_1(q^2)\gamma_\mu\gamma_5 +i g_2(q^2) \frac{\sigma_{\mu\nu}q^\nu \gamma_5}{M_A}+\frac{q_\mu g_3(q^2) \gamma_5}{M_A}\bigg] u(p_A)\nonumber,
}where $q^2=(p_A-p_B)^2$ is the squared momentum transferred in the hadronic transition, $u(p_A)$ and $M_A$ ($\bar{u}(p_B)$, and $M_B$) are the spinor and mass of the initial (final) baryon, respectively. Nevertheless,  the contributions of $f_3$, and $g_3$ form factors in Eq. (\ref{ff}) are negligible in comparison with the other form factors since they pick up a factor proportional to the mass of the charged-lepton $m_{\ell}$ involved in the transition \cite{Garcia, Schlumpf:1994fb, Ratcliffe:2004jt}. Furthermore, $f_2$ and $g_2$  are in principle not negligible, but they become subleading  in the SU(3)-flavor symmetry of QCD \cite{Sirlin:1979hb, Ratcliffe:1995fk}. Therefore, in the following we will consider that the hadronic current describing the hadronic transition in Eq. (\ref{Hmu}) is dominated by the vector and axial form factors as follows: 
\eq{
\langle B_B(p_B) \vert J_\mu  \vert B_A(p_A)\rangle= \bar{u}(p_B) \gamma_\mu \big[f_1(q^2)+ g_1(q^2) \gamma_5 \big]u(p_A).\label{Jmusimplified}
}

Now, from neutrino and electron scattering off nucleons it has been found that the observed distributions can be described by a dipole parametrization. In such a way that an extrapolation to the time-like region leads to 
\eq{
f_1(q^2)&=f_1(0)\left(1-\frac{q^2}{m_{df}^2}\right)^{-2},\\ 
g_1(q^2)&=g_1(0)\left(1-\frac{q^2}{m_{dg}^2} \right)^{-2},\label{type-pole-aprox}
}
with $ m_{df} =0.84$ GeV and $ m_{dg} =  1.08$ GeV. Since these pole masses corresponds to strangeness-conserving form factors, a rescaling using  the values of vector and axial mesons masses allows to assume that $ m_{df} =0.97$ GeV and $ m_{dg} =  1.25$ GeV  would be a good guess for the dipole masses in the strangeness-changing case \cite{Garcia, Ratcliffe:2004jt}. The values of the form factors at zero momentum transfer, $f_1(0)$ and $g_1(0)$ are given in Table \ref{t-ff} and in the case of the vector form factors they incorporate the effects of SU(3) flavor symmetry breaking \cite{Garcia, Schlumpf:1994fb, Ratcliffe:2004jt, Mateu:2005wi}.

\begin{table}[]
\begin{center}
\begin{tabular}{ c c c }
\hline \hline
Transition  &  $f_{1} (0)$  & $g_1(0)$  \\ \hline
$\Sigma^-\to n$ & -1 & 0.341   \\ 
$\Xi^-\to \Lambda$ & $\sqrt{3/2}$ & 0.239  \\ 
$\Lambda\to p$ & $-\sqrt{3/2}$ &  -0.895\\ 
\hline \hline
\end{tabular}
\end{center}
\caption{\small Vector and axial transition form factors for weak
hyperon decays at zero momentum transfer ($q^2=0$) \cite{Garcia}. }
\label{t-ff}
\end{table}

It is important to note that the amplitude $\mathcal{M}_1$ in Eq. (\ref{amp-com-1})  has a resonant effect when $(p_\pi+p_2)^2\approx m_N^2$ \footnote{
For the $B_A(p_A)\to B_B(p_B)\ell_1^-(p_1)\ell_2^-(p_2)\pi^+(p_\pi)$ decays mediated by an intermediate neutrino state produced on-shell its mass must satisfies that $m_{\ell_2^-}+m_{\pi^+}\leq m_N \leq m_A-m_B-m_{\ell_1^-}$.}. Besides, if the experiment is unable to distinguish which lepton was emitted at each stage for non-identical charged leptons or, for the antisymmetrization of identical leptons we also need to consider the diagram contribution with the final charged leptons interchanged $\ell_1(p_1) \leftrightarrow \ell_2(p_2)$ in Fig. \ref{diagram}. This second diagram has a resonant effect when $(p_\pi+p_1)^2\approx m_N^2$. Since in general $(p_\pi+p_2)^2\neq (p_\pi+p_1)^2$, it  turns out convenient to apply the Single-Diagram-Enhanced multi-channel integration method \cite{Maltoni:2002qb}. This method has been implemented for three-body channels. Here we generalize it to four-body decays, along the same lines by defining the functions 
\eq{
f_{{PS}_1}=\frac{\overline{\vert \mathcal{M}_1 \vert}^2}{\overline{\vert \mathcal{M}_1 \vert}^2+\overline{\vert \mathcal{M}_2 \vert}^2}\overline{\vert \mathcal{M} \vert}^2, \quad 
f_{{PS}_2}=\frac{\overline{\vert \mathcal{M}_2 \vert}^2}{\overline{\vert \mathcal{M}_1 \vert}^2+\overline{\vert \mathcal{M}_2 \vert}^2}\overline{\vert \mathcal{M} \vert}^2, \label{fPS-functions}
}with $\mathcal{M}=\mathcal{M}_1+\mathcal{M}_2$. In this way, Eq. (\ref{Total-amp}) can be rewritten as $\overline{\vert\mathcal{M}\vert}^2=f_{{PS}_1}+f_{{PS}_2}$, and consequently the decay width is given by 
\eq{
\Gamma_{B_A\to B_B\ell_1^-\ell_2^-\pi^+}=&\frac{N}{4(4\pi)^6m_A^3}\left[\int f_{{PS}_1} dPS_1+\int f_{{PS}_2} dPS_2\right],\label{N-width}
}with $N=1/2, (1)$ for the case where the two charged final leptons are same (different) particles. In our case, the functions $ f_{{PS}_1}$ and $f_{{PS}_2}$ can be written as follows (see the Appendix \ref{Ap-AS} for details)
\eq{
f_{{PS}_1}&=\frac{(G\,V_{\ell_1 N}V_{\ell_2 N}f_\pi m_N)^2 A}{a_1^2+\Gamma_N^2 m_N^2}\bigg[1+ 2\frac{(a_1a_2+\Gamma_N^2 m_N^2)C_1+(a_2- a_1)\Gamma_N m_NC_2}{(a_2^2+\Gamma_N^2 m_N^2)A+(a_1^2+\Gamma_N^2 m_N^2)B}\bigg],\label{int-1}\\
f_{{PS}_2}&=f_{{PS}_1}(p_1 \leftrightarrow p_2),\label{int-2}
}where the  $A$, $B$, $C_1$, and $C_2$ functions are reported for the first time in the Appendix \ref{Ap-AS}.
Now, the phase space integration can be done for $f_{{PS}_1}$ and $f_{{PS}_2}$ separately, and added up after the proper phase space integration. 
Regarding the first integral in Eq. (\ref{N-width}) \footnote{The phase space variables for the second integral in Eq. (\ref{N-width}) are chosen conveniently as ($s_{B2},\, s_{1\pi},\, \theta_B,\,  \theta_1,\, \phi$) with the mass invariants $s_{B2}=(p_B+p_2)^2$, and $s_{1\pi}=(p_1+p_\pi)^2$.}, this is described conveniently in terms of the five independent variables ($s_{B1},\, s_{2\pi},\, \theta_B,\,  \theta_2,\, \phi$)  (see Fig. 1 in reference \cite{Cabibbo:1965zzb}):
\begin{itemize}
\item $s_{B1}=(p_B+p_1)^2$ and $s_{2\pi}=(p_2+p_\pi)^2$ stand for the invariant masses of the $B_B\ell_1^-$ and $\ell_2^- \pi^+$ systems, respectively.
\item $\theta_B$ ($\theta_2$) is the angle between the three-momentum of $B_B$ ($\pi^+$) in the rest frame of the pair $B_B\ell_1^-$ ($\pi^+\ell_2^-$) with respect to the line of flight of the $B_B\ell_1^-$ ($\pi^+\ell_2^-$) in the  rest frame of the particle $B_A$.
\item $\phi$ is the angle between the planes defined by the $B_B\ell_1^-$ and $\pi^+\ell_2^-$ pairs systems in the  rest frame of the particle $B_A$.
\end{itemize} 

In order to evaluate Eq. (\ref{N-width}) we need to consider the total decay width for the new heavy neutrino states.  This can be obtained by adding up the contributions of all its partial decay widths $(\Gamma_i^{\textrm{p.w.}})$ that can be opened at the mass $m_N$ \cite{Atre:2009rg}
\eq{
\Gamma_N=\sum_{i} \Gamma_i^{\textrm{p.w.}}\cdot \theta(m_N-\sum_j m_{j}),\label{Total-width} 
}where $\theta$ is the Heaviside function and $m_{j}$ stand for the masses of all the final states particles involved in $\Gamma_i^{\textrm{p.w.}}$. Let us illustrate this point by considering  the $\Sigma^-\to n\pi^+e^-e^-$  channel, here, the mass of the resonant state must be inside the range $m_{e^-}+m_{\pi^+}\leq m_N \leq m_{\Sigma^-}-m_n-m_{e^-}$, then the possible decay channels of the heavy $N$ state (induced by charged and neutral currents) that contribute to its total decay width $\Gamma_N$ are the following  $N\to \ell^{\pm}\pi^{\mp}$, $N\to \pi^0\nu_\ell$, $N\to \ell_1^{\mp}\ell_2^{\pm}\nu_{\ell_2}$, $N\to \ell_2^{-}\ell_2^{+}\nu_{\ell_1}$, and $\nu_{\ell_1} \nu\bar{\nu}$ (with $\ell,\,\ell_1,\, \ell_2=e,\,\mu.$). The analytical expressions for these partial widths can be found in Ref. \cite{Atre:2009rg}, they depend on each particular channel considered, and they are given as a function of both the neutrino mass and the norm of the squared mixings involved, that is $\Gamma_i^\textrm{p.w}= \Gamma_i^\textrm{p.w}(m_N, \vert V_{\ell N} \vert^2)$. Then we have considered the indirect limits on the mixing elements of the heavy neutrino with the three charged leptons \cite{Fernandez-Martinez:2016lgt} in order to estimate the total neutrino width, namely 
\eq{
\vert V_{eN}\vert \leq 0.050, \quad \vert V_{\mu N}\vert \leq 0.021, \quad \vert V_{\tau N} \vert \leq 0.075.\label{mixing-ind-limits}
}Using the above values in Eq. (\ref{Total-width}) the total decay width $\Gamma_N$ varies from 0.07 neV to 4.4 neV into the resonant mass region for the $\Sigma^-\to n\pi^+e^-e^-$  decay. 
The decay width is very small compare with mass of the new neutral state $\Gamma_N \ll m_N$, and since $(p_2+p_\pi)^2=s_{2\pi}\approx m_N^2$ in Eq. (\ref{int-1}), the narrow width approximation can be applied. That means, that we can replace
\eq{
\frac{1}{(s_{2\pi}-m_N^2)^2+m_N^2\Gamma_N^2}\rightarrow\frac{\pi}{m_N\Gamma_N}\delta(s_{2\pi}-m_N^2)
}
transforming the five-variable integral in Eq. (\ref{int-1}) into a four-variable one:
\eq{
\int f_{{PS}_1} dPS_1&=\frac{\pi (G\,V_{\ell_1 N}V_{\ell_2 N}f_\pi m_N)^2}{\Gamma_N m_N}\int X\beta_{B1}\beta_{2\pi}\nonumber\\
\times &
\left[ A \left(1+2\frac{\Gamma_N^2 m_N^2 C_1+\Gamma_N m_N a_2C_2}{(a_2^2+\Gamma_N^2m_N^2)A+\Gamma_N^2m_N^2 B}\right) \right]d{s_{B1}}\,d{\cos\theta_B}\,d{\cos\theta_2}\,d\phi, 
}with the following integration limits:
\eq{
(m_B+m_1)^2&\leq s_{B1} \leq(m_A-m_2-m_\pi)^2,\quad -1\leq \cos{\theta_B}\leq 1,\nonumber\\ -1&\leq \cos{\theta_2}\leq 1,\quad\quad\quad\quad\quad\quad\quad\,\, -\pi\leq \phi \leq \pi.}
This provides all the formalism we need to compute the decay width and set the region of the parameters, given on the expected experimental branching ratio, as we show below.

\section{Numerical Analysis}

\begin{figure}
\begin{center}
\includegraphics[scale=1.2]{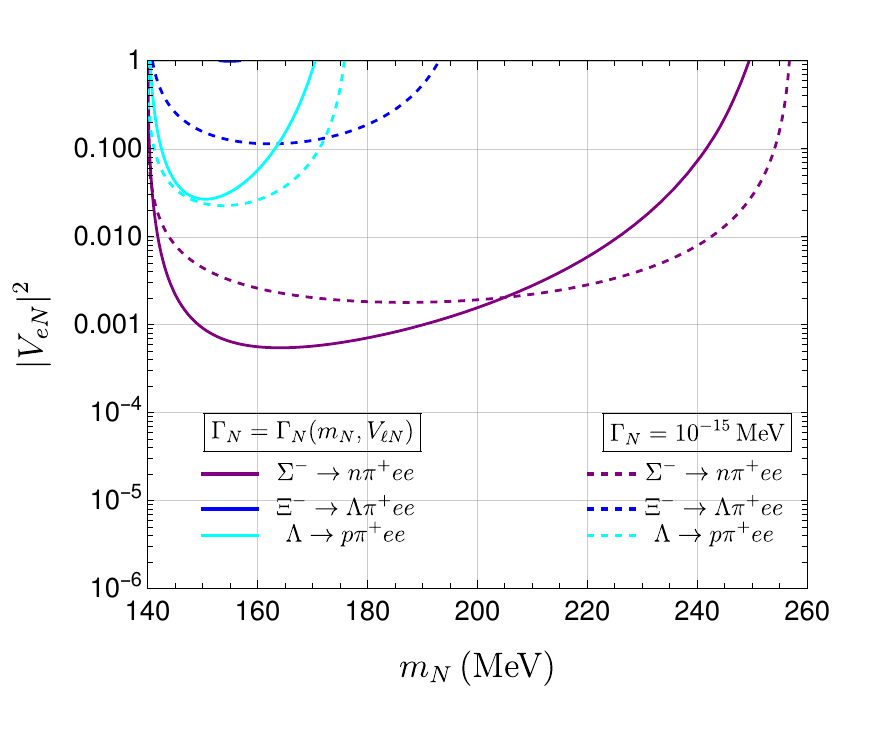}
\caption{Exclusion regions on the ($m_N, \vert V_{e N} \vert^2$) plane  by assuming a $\textrm{BR}(B_A\to B_B\pi^+e^-e^-)<10^{-8}$ limit. The purple line stand for the $\Sigma^-\to \pi^+ne^-e^-$ channel, the blue one for the $\Xi^-\to \Lambda\pi^+e^-e^-$,  and the cyan color for the $\textrm{BR}(\Lambda\to p\pi^+e^-e^-)$ decay (see main text for further details). The solid (dashed) lines correspond to estimates considering a parameters dependent neutrino width $\Gamma_N$ (fixed).}
\label{Exclusion-ee}
\end{center}
\end{figure}

\begin{figure}
\begin{center}
\includegraphics[scale=1.2]{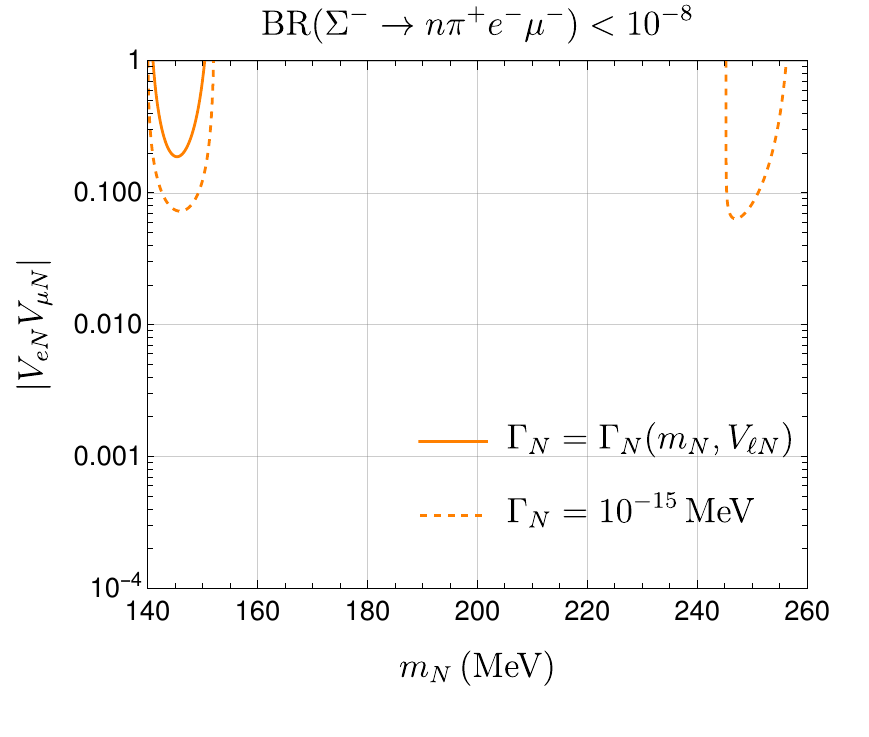}
\caption{Exclusion regions on the ($m_N, \vert V_{e N} V_{\mu N} \vert$) plane by assuming the $\textrm{BR}(\Sigma^-\to n\pi^+e^-\mu^-)<10^{-8}$ limit in the search of lepton flavor violating hyperon decays.}
\label{Exclusion-emu}
\end{center}
\end{figure}

The projected sensitivity of BES-III for the search of rare and forbidden hyperon three-body hyperon decays at BES-III is of the order of $10^{-6}-10^{-8}$  \cite{Li:2016tlt} with clean backgrounds \footnote{It is also worthy to mention that these kinds of transitions can be also searched by the LHCb collaboration with higher sensitivities because of the huge production cross-section there.
}. However, there is not an estimation for similar four-body hyperon decays. In this work, we will assume an optimistic scenario considering similar sensitivities for three and four-body processes. 

In Fig. \ref{Exclusion-ee} we show the exclusion region on the plane ($m_N$, $\vert V_{eN}\vert^2$) for the neutrino resonant state obtained by assuming a rate of BR$(B_A \to B_B\pi^+ e^-e^-)<10^{-8}$ for the channels involving a pair of electrons in the final state. We have considered here two benchmarks to evaluate the total neutrino width. On one side, the solid lines represent the universal coupling assumption, that is, we consider that $V_{eN}=V_{\mu N}=V_{\tau N}$ in Eq. (\ref{Total-width}); therefore, the total neutrino width (and consequently the branching ratio of the $B_A \to B_B\pi^+ e^-e^-$ hyperon decays) can be expressed only as a function of $\Gamma_N=\Gamma_N(\vert V_{eN} \vert^2, m_N)$. On the other hand, the dashed lines represent a scenario where the total neutrino width is fixed to the razonable value $\Gamma_N=10^{-15}$ MeV (consistent with the estimation of the total neutrino width using the indirect limits reported in Eq. (\ref{mixing-ind-limits})).  From this plot, we can observe that, in general, the exclusion region will depend on which assumption we considered, although in general, they are of the same order for all the allowed mass of the resonant neutrino in the different channels. In any case, the most restrictive limits  will come from the $\Sigma^-\to n\pi^+e^-e^-$ channel, follow by $\Lambda\to p\pi^+e^-e^-$, and finally the much less restrictive $\Xi^-\to \Lambda\pi^+e^-e^-$ channel. 
Additionally to the processes with a pair of electrons in the final state, the $\Sigma^-\to n\pi^+e\mu^-$ channel is the only possible kinematically allowed four-body LNV hyperon decay. As we can see in Fig. \ref{Exclusion-emu}, if the search for this transition can achieve a rate of BR$(\Sigma^- \to n\pi^+ e^-\mu^-)<10^{-8}$ then the limits set on the plane ($m_N, \vert V_{eN} V_{\mu N}\vert$) are much less restrictive than the di-electronic case because phase space restrictions are more stringent. For the case with two different flavors notice that the limits are split into two disconnected regions. The left (right) region on Fig \ref{Exclusion-emu}, is associated with the case where the muon (electron) was created along with the resonant neutrino state, and the electron (muon) comes after the subsequent neutrino decay. Overlap of these regions can be achieved in other scenarios, provided the kinematical conditions allow them to do so. The formalism here developed allow to address both cases regardless of invoking the direct narrow width approximation or not (see appendix \ref{dnwa}).

\section{Conclusions}

The search for $\Delta$L=2 processes is crucial for unraveling the Dirac or Majorana nature of neutrinos.  Except possibly for neutrinoless double-beta decay in nuclei, diverse neutrino mass models predict that LNV effects can lie beyond the reach of current experiments. However, if new hypothetical heavy Majorana neutrinos with masses from $\sim 100$ MeV to few GeV can be produced on-shell as an intermediate state in LNV decays of mesons, baryons, or the tau lepton, then their branching ratios can be amplified due to a resonant effect. The no observation of such processes sets limits on the parameter space of these new heavy neutrinos states. In this regard, most of the studies have focused on three-body LNV meson or tau decays, however, recently the study of similar four-body LNV channels has also drawn attention as complementary means because they can provide information about different kinematical phase-space regions. In this work, we studied the four-body LNV decays of hyperons mediated by a resonant Majorana neutrino.

Our results suggest that the direct limits derived on $\vert V_{eN} \vert^2$ from the $\Sigma^-\to n\pi^+e^-e^-$ channel can be of the same order ($\sim 10^{-3}$) than those obtained from the meson decays, such as $D^+\to \pi^-e^+e^+$ and $D_s^+\to \pi^-e^+e^+$, but far away from the current most stringent ones from the  semileptonic kaon decay $K^+\to \pi^{-} e^+e^+$  which is around $\mathcal{O}(10^{-9})$ \cite{Abada:2017jjx} \footnote{Current bounds for $K^+\to \pi^{-} e^+e^+ \leq 5.3 \times 10^{-11}$ and $K^+\to \pi^{-} e^+\mu^+ \leq 4.2 \times 10^{-11}$  are reported by the NA62 experiment at CERN in \cite{NA62:2022tte, NA62:2019eax}.}. Moreover, less restrictive limits on $\vert V_{eN} \vert^2$ ($\sim 10^{-1}$) can be obtained from the $\Lambda\to p\pi^+e^-e^-$ and  $\Xi^-\to \Lambda\pi^+e^-e^-$ which are comparable with the limits from the $B^+\to \pi^-e^+e^+$ meson channel. On the other hand, the $\Sigma^-\to n\pi^+e^-\mu^-$ channel is the only possible four-body LNV hyperon decay mediated by a Majorana neutrino involving a muon as final state, but places a very weak contraints on $\vert V_{eN}V_{\mu N} \vert$ for the small mass on-shell neutrino regions allowed assuming the expected sensitivity for rare hyperon decays of BES III.


\section*{Acknowledgements}
We would like to thank G. L\'opez Castro for many helpful discussions.
G.H.T. and G.T. acknowledge the support of DGAPA-PAPIIT UNAM, grant no. IN110622 for financial support. G.H.T. is funded by PROGRAMA DE BECAS POSDOCTORALES DGAPA-UNAM. 
The work of D.P.S. was supported by Ciencia de Frontera Conacyt project No. 428218 and the program ``BECAS CONACYT NACIONALES''.

\appendix

\section{Appendix}\label{Ap-AS}
Following the Feynman rules for fermion number-violating interactions reported in Ref. \cite{Denner:1992vza}, the contribution of the second diagram with the charged leptons interchanged in Fig. \ref{diagram}(a) is given by
\eq{
\mathcal{M}_{2}=\left(\frac{G\,V_{\ell_1 N}V_{\ell_2 N}f_\pi m_N}{a_{2}+i\Gamma_N m_N}\right)\ell_{\nu\mu}(p_{1},p_{2})\, H^\mu(p_B, p_A)\, p_\pi^\nu,\label{amp-com-2}
} 
with $a_{2}\equiv(p_A-p_B-p_{2})^2-m_N^2$. Therefore, $\mathcal{M}=\mathcal{M}_{1}+\mathcal{M}_{2}$ can be written as follows
\eq{
\mathcal{M}=G\,V_{\ell_1 N}V_{\ell_2 N} f_\pi m_N \bar{u}(p_1) \left(\frac{\gamma_\mu \gamma_\nu}{{a_{1}+i\Gamma_N m_N}}+\frac{\gamma_\nu \gamma_\mu}{{a_{2}+i\Gamma_N m_N}}\right)(1+\gamma_5)v(p_2)\, H^\mu(p_B, p_A)\, p_\pi^\nu.
}
The total amplitude squared is given by
\eq{
\overline{\vert\mathcal{M}\vert}^2 &=\frac{1}{2}\sum_{\textrm{spins}}\vert \mathcal{M}\vert^2=\frac{1}{2}\sum_{\textrm{spins}}\left(\vert\mathcal{M}_1\vert^2+\vert\mathcal{M}_2\vert^2 + 2\textrm{Re}[\mathcal{M}_1\mathcal{M}_2^{\dagger}]\right),\label{Total-amp}
}where the individual contributions can be written as follows 
\eq{
\overline{\vert\mathcal{M}_1\vert}^2&=\frac{1}{2}\sum_{\textrm{spins}}\vert\mathcal{M}_1\vert^2=\frac{1}{2}\frac{(G\,V_{\ell_1 N}V_{\ell_2 N}f_\pi m_N)^2 A}{\left(a_1^2+\Gamma_N^2 m_N^2\right)}\label{ind-ms}\\
\overline{\vert\mathcal{M}_2\vert}^2&=\frac{1}{2}\sum_{\textrm{spins}}\vert\mathcal{M}_2\vert^2=\frac{1}{2}\frac{(G\,V_{\ell_1 N}V_{\ell_2 N}f_\pi m_N)^2 B}{\left(a_2^2+\Gamma_N^2 m_N^2\right)}\nonumber,
}
while the interference term
\eq{
\mathcal{M}_1\mathcal{M}_2^\dagger&=(G\,V_{\ell_1 N}V_{\ell_2 N}f_\pi m_N)^2(C_1+iC_2)\frac{\left[a_1a_2+\Gamma_N^2m_N^2+i(a_1-a_2)\right]\Gamma_N m_N}{\left(a_1^2+\Gamma_N^2 m_N^2\right)\left(a_2^2+\Gamma_N^2 m_N^2\right)},
}with the $A,$ $B,$ $C_1$, and $C_2$ functions given by:
\eq{
A&=64[f_1^2(q^2)\xi_1+g_1^2(q^2)\xi_2+f_1(q^2)g_1(q^2)\xi_3],\\
B&=A(p_1\leftrightarrow p_2), \\
C_1&=64[f_1^2(q^2)\xi_4+g_1^2(q^2)\xi_5+f_1(q^2)g_1(q^2)\xi_6]\nonumber\\C_2&=64\epsilon_{\mu\nu\lambda\rho}p_B^\mu p_1^\nu p_2^\lambda p_\pi^\rho[-(f_1^2+g_2^2)(r_{A\pi}+r_{B\pi})+f_1(q^2)g_1(q^2)2(r_{1\pi}+r_{2\pi})],
}and the following definitions
\eq{
\xi_1&=m_Am_B(m_\pi^2r_{12}-2r_{1\pi} r_{2\pi})-m_\pi^2(r_{A1}r_{B2}+r_{A2}r_{B1})\label{xi1}+2r_{2\pi}(r_{A1}r_{B\pi}+2r_{A\pi}r_{B1}),\\
\xi_2&=\xi_1-2m_Am_B(m_\pi^2 r_{12}-2r_{12}r_{2\pi}),\label{xi2}\\
\xi_3&=2[m_\pi^2(r_{A2}r_{B1}-r_{A1}r_{B2})+2r_{2\pi}(r_{A1}r_{B\pi}-r_{A\pi}r_{B1})],\label{xi3}\\
\xi_4&=-2m_Am_B r_{1\pi}r_{2\pi}+m_\pi^2(r_{12}r_{AB}-r_{A1}r_{B2}-r_{A2}r_{B1})-2r_{12}r_{A\pi}r_{B\pi}\label{xi4}\nonumber\\
&+r_{1\pi}(r_{A2}r_{B\pi}+r_{A\pi}r_{B2})+r_{2\pi}(r_{A1}r_{B\pi}+r_{A\pi}r_{B1})\\
\xi_5&=\xi_4-2m_Am_B(m_\pi^2 r_{12}-2r_{1\pi}r_{2\pi}),\label{xi5}\\
\xi_6&=2[r_{1\pi}(r_{A2}r_{B\pi}-r_{A\pi}r_{B2})+r_{2\pi}(r_{A1}r_{B\pi}-r_{A\pi}r_{B1})].\label{xi6}
}In the above expression, we have defined $r_{ij}\equiv p_i\cdot p_j$ with $p_i$ and $p_j$ denoting any of the momenta of the external particles  (that is $p_{i,j}=p_A,p_B,p_1,p_2,p_\pi$).  Now, for the set of variables chosen in Section \ref{S-com},  the scalar products $r_{ij}$ involved in Eqs. (\ref{xi1}-\ref{xi6}) are given as follows
\eq{
r_{B1}&=\frac{1}{2}(s_{B1}-m_B^2-m_1^2),\quad\quad
r_{2\pi}=\frac{1}{2}(s_{2\pi}-m_2^2-m_\pi^2),\\
r_{B2}&=\frac{1}{4}(\alpha_1+\alpha_2+\alpha_3+\alpha_4),\quad
r_{B\pi}=\frac{1}{4}(\alpha_1-\alpha_2+\alpha_3-\alpha_4),\\
r_{12}&=\frac{1}{4}(\alpha_1+\alpha_2-\alpha_3-\alpha_4),\quad
r_{1\pi}=\frac{1}{4}(\alpha_1-\alpha_2-\alpha_3+\alpha_4),\\
r_{AB}&=\frac{1}{2}(\alpha_5+\alpha_6),\quad\quad\quad\quad\quad
r_{A1}=\frac{1}{2}(\alpha_5-\alpha_6),\\
r_{A2}&=\frac{1}{2}(\alpha_7+\alpha_8),\quad\quad\quad\quad\quad
r_{A\pi}=\frac{1}{2}(\alpha_7-\alpha_8),\\
\epsilon_{\mu\nu\lambda\rho}p_B^\mu p_1^\nu p_2^\lambda p_\pi^\rho&=-\sqrt{s_{B1}s_{2\pi}}\beta_{B1}\beta_{2\pi}X\sin{\theta_B}\sin{\theta_2}\sin{\phi},
}with the definitions
\eq{
\alpha_1&=\frac{1}{2}(m_A^2-s_{B1}-s_{2\pi}),\\
\alpha_2&=X\beta_{2\pi}\cos{\theta_2}+\left(\frac{m_2^2-m_\pi^2}{s_{2\pi}} \right)\alpha_1,\\
\alpha_3&=X\beta_{B1}\cos{\theta_B}+\left(\frac{m_B^2-m_1^2}{s_{B1}} \right)\alpha_1,\\
\alpha_4&=\left(\frac{m_B^2-m_1^2}{s_{B1}} \right)\left(\frac{m_2^2-m_\pi^2}{s_{2\pi}} \right)\alpha_1+\left(\frac{m_B^2-m_1^2}{s_{B1}} \right)X\beta_{2\pi}\cos{\theta_2}\\
&+\left(\frac{m_2^2-m_\pi^2}{s_{2\pi}} \right)X\beta_{B1}\cos{\theta_B}+\beta_{B1}\beta_{2\pi}\left(\alpha_1\cos{\theta_B}\cos_{\theta_2}-\sqrt{s_{B1}s_{2\pi}}\sin{\theta_B}\sin{\theta_2}\cos{\phi}\right)\nonumber\\
\alpha_5&=\frac{1}{2}(m_A^2+s_{B1}-s_{2\pi}),\\
\alpha_6&=(m_B^2-m_1^2)\left(1+\frac{\alpha_1}{s_{B1}}\right)+X\beta_{B1}\cos{\theta_B},\\
\alpha_7&=\frac{1}{2}(m_A^2-s_{B1}+s_{2\pi}),\\
\alpha_8&=(m_2^2-m_\pi^2)\left(1+\frac{\alpha_1}{s_{2\pi}}\right)+X\beta_{2\pi}\cos{\theta_2},
}and
\eq{
\lambda(a,b,c)&= a^2+b^2+c^2-2(ab+bc+ac),\\
X&=\frac{\lambda(m_A^2,s_{B1},s_{2\pi})^{1/2}}{2},\quad
\beta_{B1}=\frac{\lambda(s_{B1},m_B^2,m_1^2)^{1/2}}{s_{B1}},\quad
\beta_{2\pi}=\frac{\lambda(s_{2\pi},m_2^2,m_\pi^2)^{1/2}}{s_{2\pi}}.
}

\section{Comparison with direct narrow width approximation computation}\label{dnwa}
For completeness and as a crosscheck of our computation we have verified that for the cases where we can distinguish the flavour of the charged lepton created as a product of the decay of the resonant state or the channels with two identical external charged leptons, the results using the Single-Diagram-Enhanced multi-channel integration method can be reproduced by applying directly the narrow width approximation  
\eq{
\textrm{BR}(B_A\to B_B\ell_1^-\ell_2^-\pi^+)= \textrm{BR}(B_A\to B_B\ell_1^- N)\times \Gamma (N\to \ell_2^-\pi^+)\tau_N/\hbar,\label{NWA}
}where $\tau_N$ is the lifetime of the intermediate neutrino state. In this case, the partial decay width $\Gamma (N\to \ell_2^-\pi^+)$ can be computed straightforwardly by \cite{Atre:2009rg}:
\eq{
\Gamma(N\to \ell_2^-\pi^+)=\frac{G_F^2}{16\pi}\vert V_{ud} \vert ^2 \vert V_{\ell_2 N} \vert^2 f_\pi^2 m_N\, \lambda^{\frac{1}{2}}\left(m_N^2, m_{\ell_2}^2, m_\pi^2 \right)\big[\left(1-x_{\ell_2}\right)^2-x_\pi\left(1+x_{\ell_2}\right) \big],
}with $x_y\equiv m_y^2/m_N^2$, $\lambda$ is the K\"allen function defined previously, and $f_\pi$ is the pion decay constant. Regarding the subprocess $B_A(p_A)\to B_B(p_B)\ell_1^-(p_1)N(p_N)$ in Eq. (\ref{NWA}), the amplitude is given by
\eq{
\mathcal{M}(B_A\to B_B\ell_1^-N)=-\frac{G_F}{\sqrt{2}}V_{us} V_{\ell_1 N}\langle B_B(p_B) \vert J_\mu  \vert B_A(p_A)\rangle L^{\mu},\label{amp}
}with $J_\mu$ hadronic previously defined in Eq. (\ref{Jmusimplified}), and the leptonic current defined as follows
\eq{
L^\mu\equiv\bar{u}(p_1)\gamma^\mu (1-\gamma_5)v(p_N).
}
The squared amplitude of Eq.  (\ref{amp}) is given by
\eq{
\vert\mathcal{M}\vert^2 =\mathcal{M} \mathcal{M}^\dagger =& 32 G_F^2\vert V_{us}\vert^2 \vert V_{\ell_1 N} \vert^2 \bigg[m_A m_B \big(g_1^2(q^2)-f_1^2(q^2)\big)\\ \nonumber
&+\big(f_1(q^2)-g_1(q^2)\big)^2 (p_1\cdot p_B) (p_A\cdot p_N)\nonumber\\
&+\big(f_1(q^2)+g_1(q^2)\big)^2 (p_1\cdot p_A) (p_B\cdot p_N) \bigg].\nonumber
}Now, by defining $s_{1N}\equiv (p_1+p_N)^2$ and $s_{1B}\equiv (p_1+p_B)^2$, the branching ratio can be expressed in terms of these two Lorentz invariants as follows
\eq{
\textrm{BR}(B_A\to B_B\ell_1^-N)= \frac{G_F^2 \vert V_{us}\vert^2 \vert V_{\ell_1 N} \vert^2}{64\pi^3m_A^3 \Gamma_{B_A}}\int_{s_{{1B}_\textrm{min}}}^{s_{{1B}_\textrm{max}}}\int_{s_{{1N}_\textrm{min}}}^{s_{{1N}_\textrm{max}}}\mathcal{F}(s_{1B}, s_{1N})ds_{1N}ds_{1B},
}where
\eq{
\mathcal{F}(s_{1B}, s_{1N})&=2m_Am_B\big(s_{1N}-m_N^2-m_1^2\big)\bigg[\frac{g_1^2(0)}{\big(1-\frac{s_{1N}}{m_{d_g}^2}\big)^4}-\frac{f_1^2(0)}{\big(1-\frac{s_{1N}}{m_{d_f}^2}\big)^4}\bigg]\\
&+\big(s_{1B}-m_B^2-m_1^2 \big)\big(m_A^2+m_N^2-s_{1B} \big)\bigg[\frac{f_1(0)}{\big(1-\frac{s_{1N}}{m_{d_f}^2}\big)^2}-\frac{g_1(0)}{\big(1-\frac{s_{1N}}{m_{d_g}^2}\big)^2}\bigg]^2\nonumber\\
&+\big(m_1^2+m_A^2-s_{1B}-s_{1N} \big)\big(s_{1B}+s_{1N}-m_N^2-m_B^2 \big)\bigg[\frac{f_1(0)}{\big(1-\frac{s_{1N}}{m_{d_f}^2}\big)^2}+\frac{g_1(0)}{\big(1-\frac{s_{1N}}{m_{d_g}^2}\big)^2}\bigg]^2, \nonumber
}and the phase-space integration limits are
\eq{
s_{1B}^{\pm}=&m_A^2+m_1^2-s_{1N}-\frac{1}{2s_{1N}}\bigg[(m_A^2-m_B^2-s_{1N})(s_{1N}-m_1^2+m_2^2)\\
&\pm \sqrt{\lambda(m_A^2,m_B^2, s_{1N})\lambda(s_{1N},m_1^2,m_2^2)}\bigg]\nonumber,
}and
\eq{
(m_1+m_N)^2\leq s_{1N}\leq (m_A-m_B)^2.
}
Finally, the numerical values (central values) for the masses, lifetimes, and CKM elements used in our numerical analysis are reported in \cite{PDG}.

\end{document}